\documentclass[a4paper,11.5pt]{article}
\usepackage{graphicx}
\usepackage{epstopdf}
\usepackage{amssymb,amsmath}
\usepackage[colon,authoryear]{natbib}
\usepackage{multirow}
\usepackage{booktabs}
\newlength{\verticalcompensationlength}
\newcounter{verticalcompensationrows}

\title{1/1 resonant periodic orbits \\in three dimensional planetary systems}

\author{Kyriaki I. Antoniadou, George Voyatzis and Harry Varvoglis\\
\small{Section of Astrophysics, Astronomy and Mechanics}, \\ \small{Department of Physics, Aristotle University of Thessaloniki,} \\
\small{Thessaloniki, 54124, Greece}\\
\small{kyant@auth.gr, voyatzis@auth.gr and varvogli@physics.auth.gr}}

\begin{document}

\maketitle
\begin{abstract}
We study the dynamics of a two-planet system, which evolves being in a $1/1$ mean motion resonance (co-orbital motion) with non-zero mutual inclination. In particular, we examine the existence of bifurcations of periodic orbits from the planar to the spatial case. We find that such bifurcations exist only for planetary mass ratios $\rho=\frac{m_2}{m_1}<0.0205$. For $\rho$ in the interval $0<\rho<0.0205$, we compute the generated families of spatial periodic orbits and their linear stability.  These spatial families form bridges, which start and end at the same planar family. Along them the mutual planetary inclination varies. We construct maps of dynamical stability and show the existence of regions of regular orbits in phase space.
\end{abstract}

{\bf keywords} co-orbital motion, mutual inclination, spatial periodic orbits, planetary systems
\vspace{.5em}

Planar co-orbital motion for a planetary system has been studied analytically (e.g. see \cite{r13}) or numerically (e.g. \cite{h11, h12}). In the last two papers,  the main families of periodic orbits have been determined and classified as {\em planetary} or {\em satellite}. Herein, we utilize the spatial general TBP in a rotating frame of reference (\cite{av13}) and study co-orbital motion in space. 
   
\begin{figure}[H]
\begin{center}
$\begin{array}{@{\hspace{-.5em}}ccc}
\includegraphics[width=4.15cm,height=4.25cm]{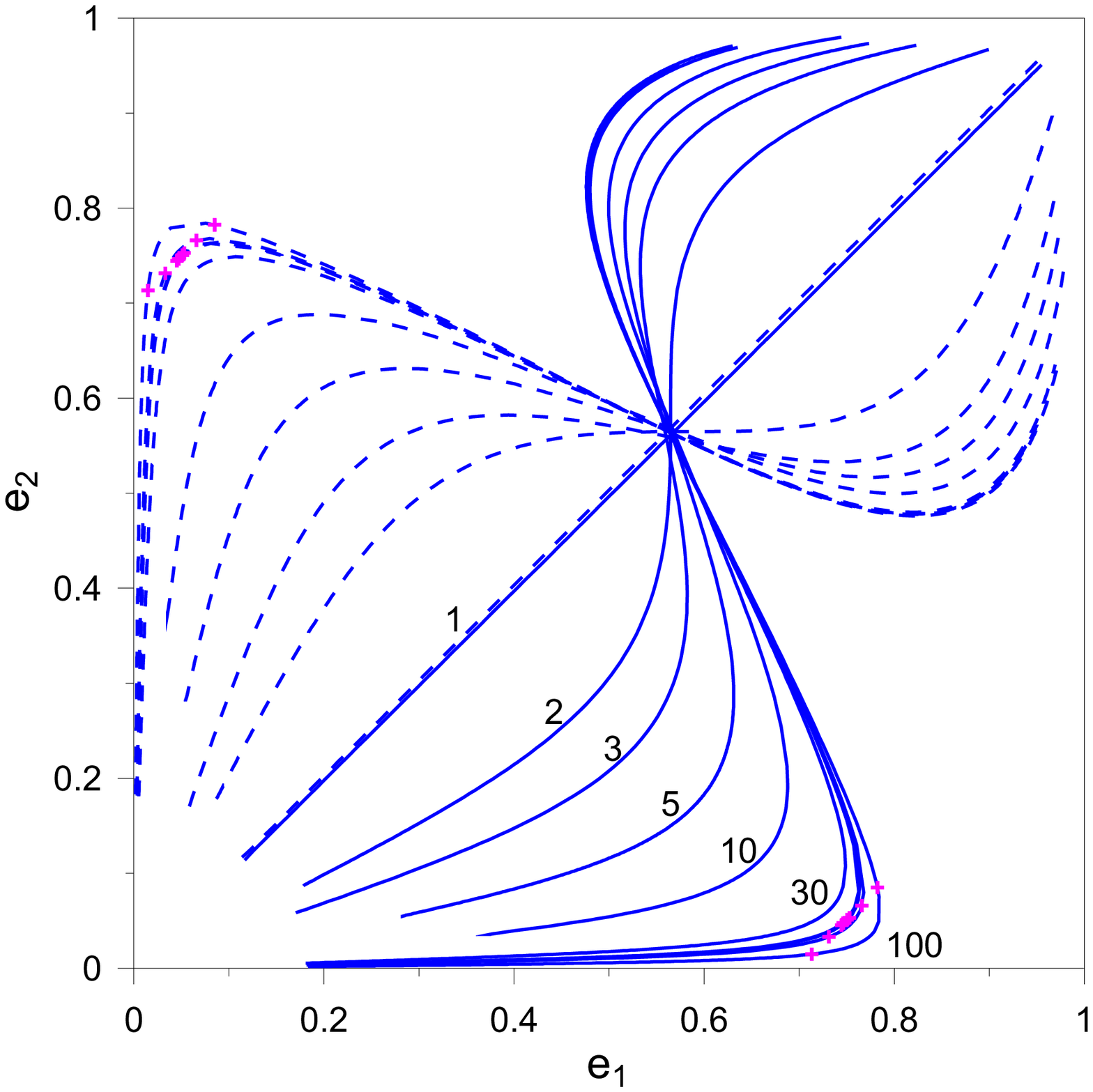}&
\includegraphics[width=4.15cm,height=4.25cm]{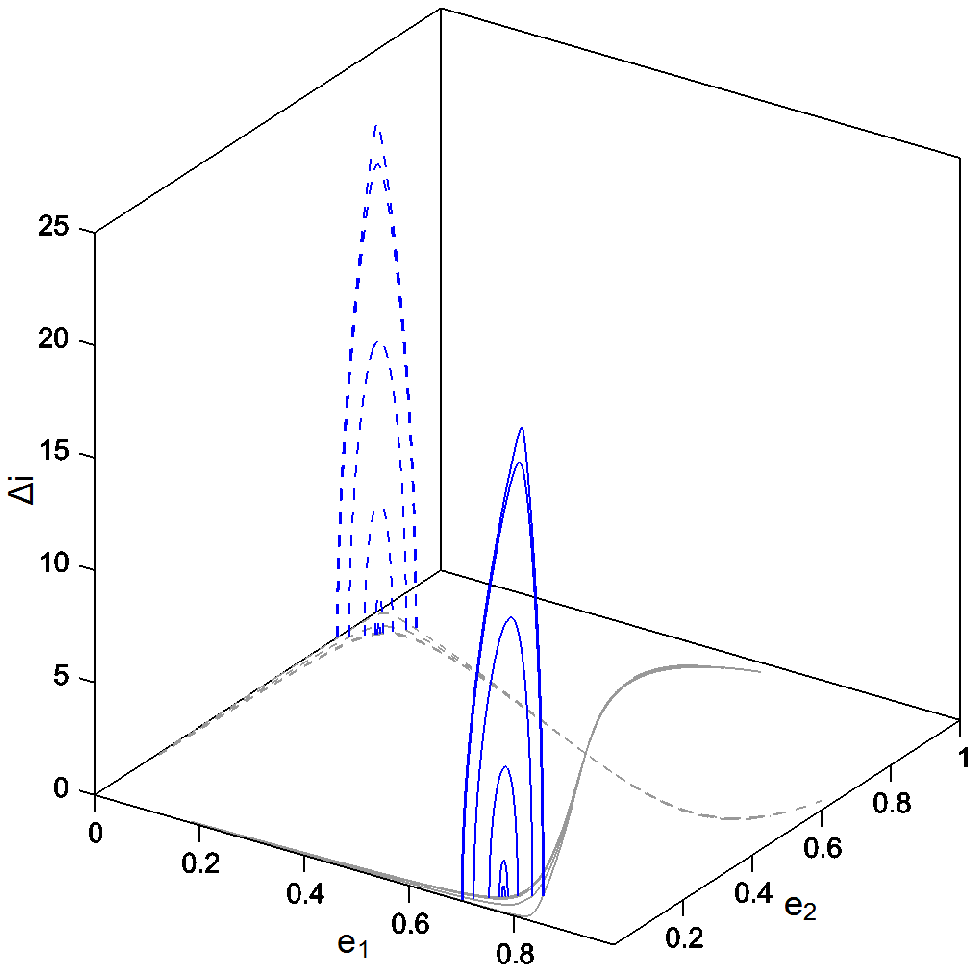}  &
\includegraphics[width=4.15cm,height=4.25cm]{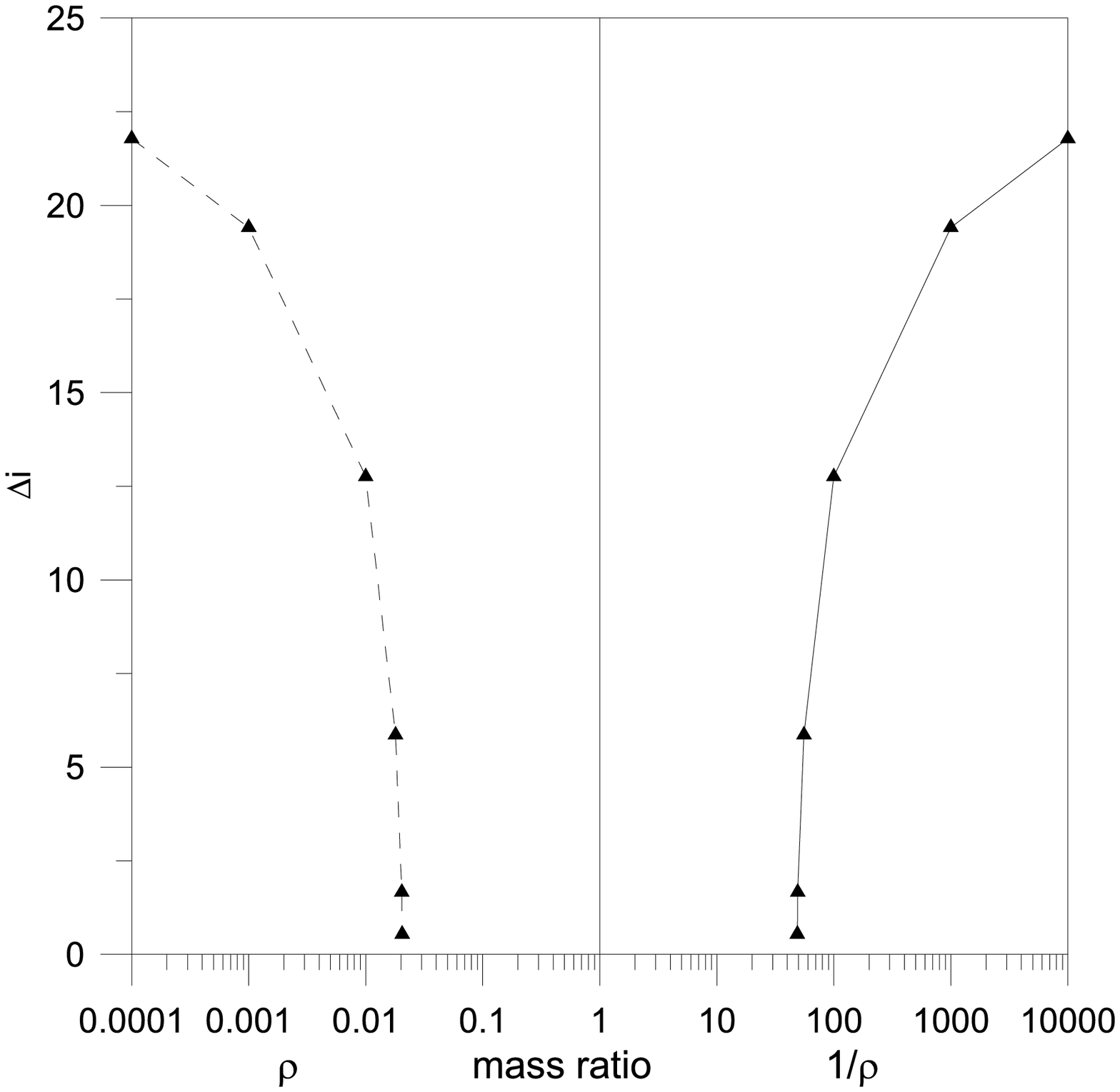}  \\
\textnormal{(a)} & \textnormal{(b)} & \textnormal{(c)} \\ 
\end{array} $
\end{center}
\caption{{\bf a} Planar and {\bf b} spatial families of symmetric periodic orbits in $1/1$ resonance. {\bf c} The maximum mutual inclination along families as a function of $\rho$. 
}
\label{fig1}
\end{figure}

Firstly, we examine the vertical stability (\cite{av13}) of all families of planar stable periodic orbits mentioned above. We find that all periodic co-planar orbits are vertically stable, which means that, if we assume small deviations from the co-planar motion, the long-term stability of the orbits will be preserved with only small anti-phase oscillations in the planetary inclinations (i.e. when $i_1$ increases, $i_2$ decreases and vice-versa, due to the invariance of the total angular momentum). However, the families of planar symmetric planetary orbits, which are projected on the eccentricity plane in Fig. \ref{fig1}a, show segments of vertical instability, if the planetary mass ratio is $\rho=m_2/m_1 \leq 0.02049$ (or $\rho\geq48.804$ if $m_1<m_2$). At the edges of vertically unstable segments we have vertical critical orbits (v.c.o) (see Table \ref{11t} and $(+)$ symbols in Fig. \ref{fig1}a).     

\begin{table}
\caption{Eccentricity values $(e_1,e_2)$ of v.c.o. in $1/1$ resonance.}
\begin{tabular}[htb]{c|cccc}
\toprule
$\rho\; (1/\rho)$ & \hspace{0.25em} 0.01 (100) & \hspace{0.25em} 0.018 (55.55) &  \hspace{0.25em} 0.02 (50) & \hspace{0.25em} 0.0205 (48.8) \\
\midrule
v.c.o. 1  & \hspace{0.25em}(0.015, 0.713) & \hspace{0.25em}(0.033, 0.731) & \hspace{0.25em}(0.045, 0.744) & \hspace{0.25em}(0.047, 0.746) \\ 
v.c.o. 2  & \hspace{0.25em}(0.085, 0.782) & \hspace{0.25em}(0.066, 0.766) & \hspace{0.25em}(0.053, 0.752) & \hspace{0.25em}(0.051, 0.750)  \\
\bottomrule\end{tabular}
\label{11t}
\end{table}
   
Families of spatial $xz$-symmetric periodic orbits bifurcate from the v.c.o. and form bridges which connect the couples of v.c.o. In Fig. \ref{fig1}b, these families are presented on the $3D$ space $e_1-e_2-\Delta i$, where $\Delta i$ denotes the mutual inclination. The spatial periodic orbits correspond to almost Keplerian orbits with $\omega_2-\omega_1=0$, $\Omega_2-\Omega_1=\pi$, where $\omega_i$ and $\Omega_i$ denote the arguments of pericenter and the longitudes of ascending node, respectively. Also, at $t=0$ one planet starts at periastron and the other at apoastron. Along the family $\Delta i$ reaches a maximum value, which depends on $\rho$ and is presented in Fig. \ref{fig1}c. All spatial periodic orbits computed are linearly stable.
   
\begin{figure}[H]
\begin{center}
$\begin{array}{@{\hspace{-.5em}}ccc}
\includegraphics[width=4.15cm,height=4.25cm]{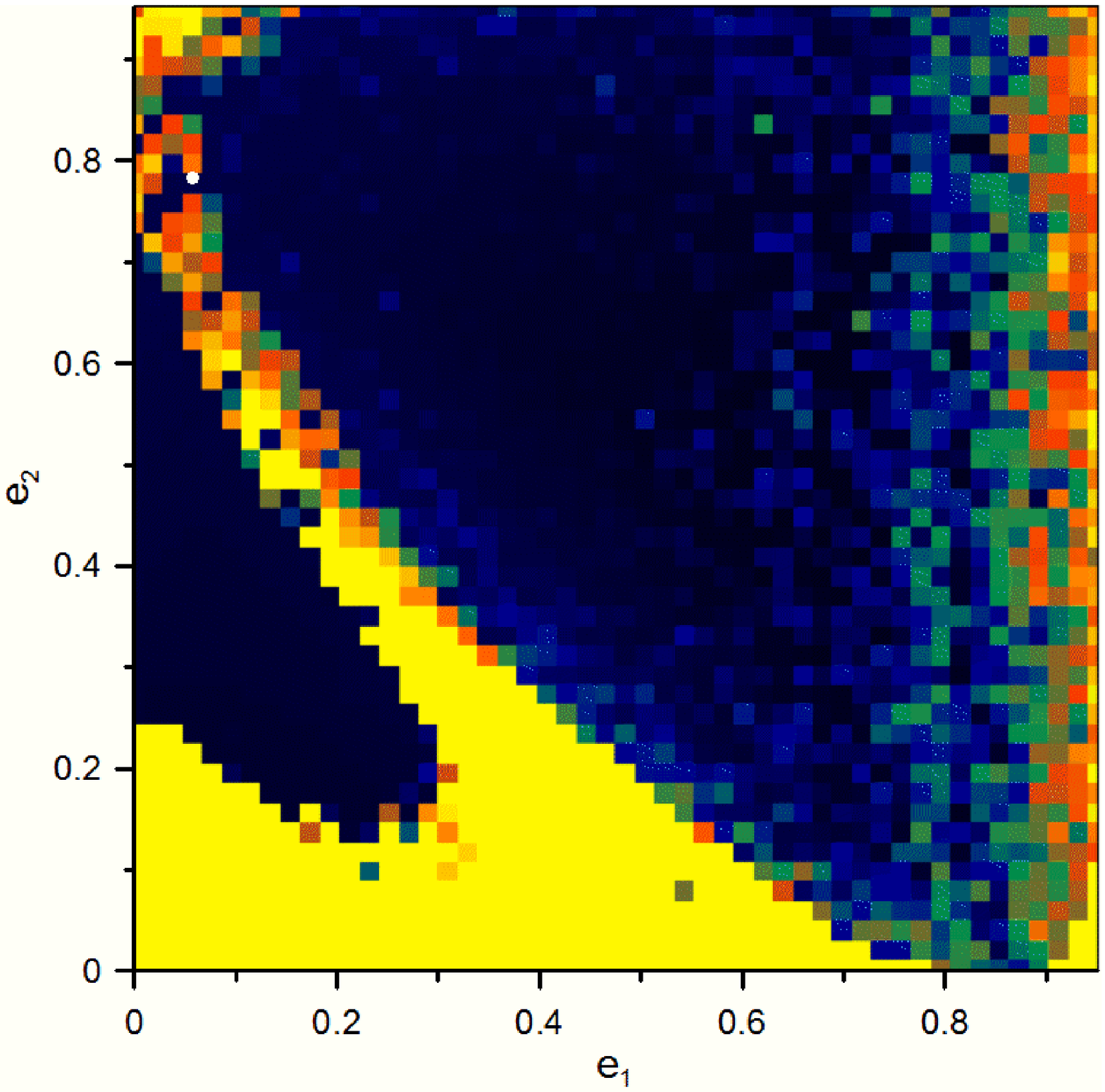}  &
\includegraphics[width=4.15cm,height=4.25cm]{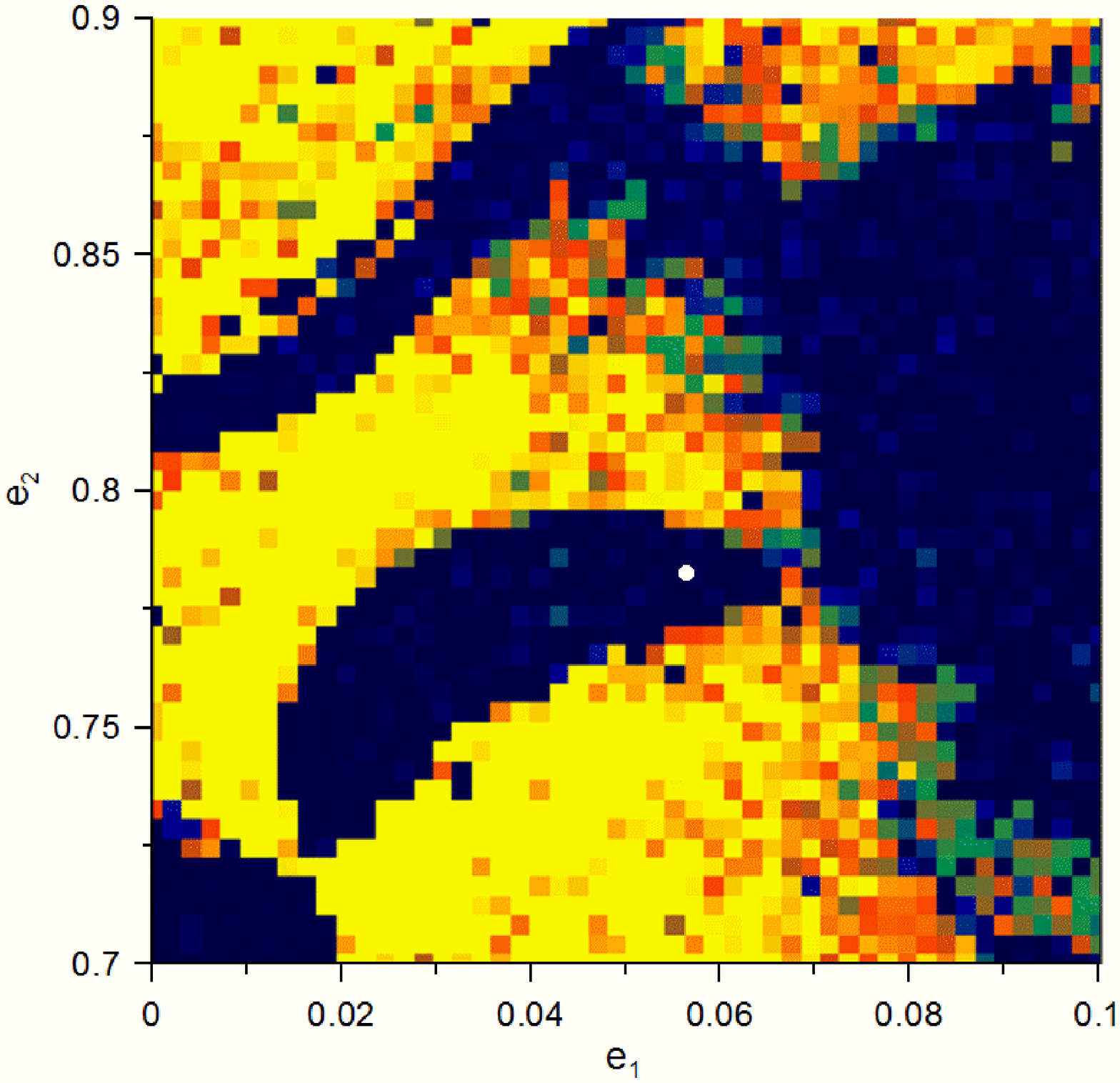}  &
\includegraphics[width=4.15cm,height=4.25cm]{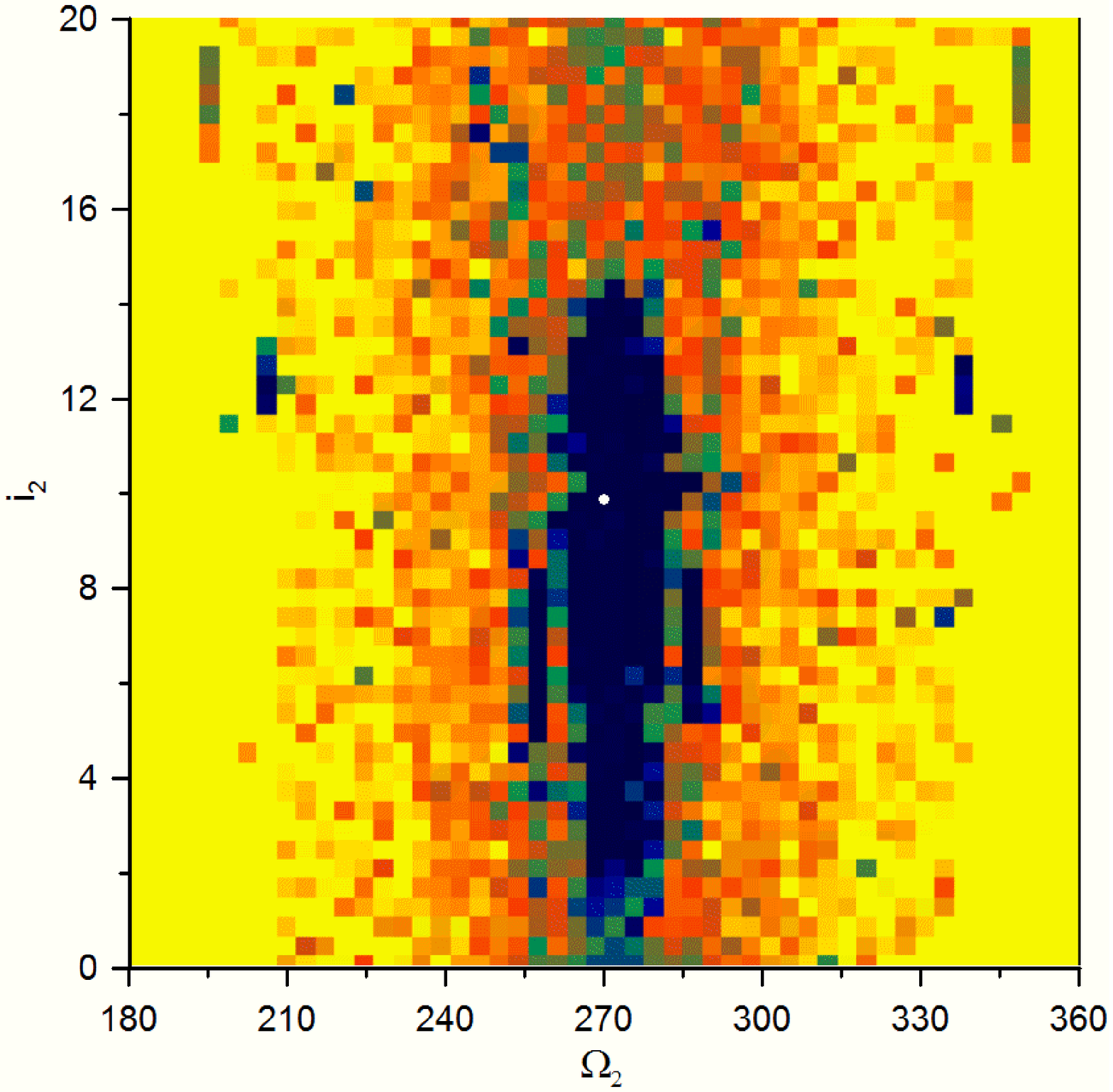}   \\
\textnormal{(a)} & \textnormal{(b)} & \textnormal{(c)} \\ 
\end{array} $
\end{center}
\caption{Maps of dynamical stability for $\rho=m_2/m_1=0.01$.} 
\label{fig2}
\end{figure}

We construct maps of dynamical stability assuming plane grids of initial conditions and computing a FLI for chaos detection (\cite{gv}). On these maps, dark colour indicates regular orbits, while regions of pale colour indicate strongly irregular orbits (Fig. \ref{fig2}). Apart from the initial conditions that correspond to the particular grid, the other ones are those of the periodic orbit with $\Delta i \approx 10^\circ$. The map of panel (a) indicates the existence of stable orbits in a large eccentricity domain (but for the particular phase configuration). Panel (b) is a magnification of the previous map around the periodic orbit, which is indicated by a white dot, and seems to define an isolated domain (island) of regular motion. In panel (c), it is clearly shown that stability is restricted for $\Omega_2\approx 270^\circ$ and consequently, for  $\Omega_2-\Omega_1\approx 180^\circ$.    

\vspace{.5em}
{\bf Acknowledgments:} This research has been co-financed by the European Union (European Social Fund - ESF) and Greek national funds through the Operational Program ``Education and Lifelong Learning'' of the National Strategic Reference Framework (NSRF) - Research Funding Program: Thales. Investing in knowledge society through the European Social Fund.


\begin{thebibliography}{}

\bibitem[Antoniadou and Voyatzis (2014)]{av13}
{Antoniadou, K.~I., \& Voyatzis, G.} 2014,
\textit{Astrophys. Space Sci.}, 349, 657

\bibitem[Hadjidemetriou et al. (2009)]{h11}
{Hadjidemetriou, J.~D., Psychoyos, D. \& Voyatzis, G.} 2009, 
\textit{Celest. Mech. Dyn. Astr.}, 104, 23

\bibitem[Hadjidemetriou and Voyatzis (2011)]{h12}
{Hadjidemetriou, J.~D. \& Voyatzis, G.} 2011, 
\textit{Celest. Mech. Dyn. Astr.}, 111, 179

\bibitem[Robutel and Pousse (2013)]{r13}
{Robutel, P. \& Pousse, A.} 2013, 
\textit{Celest. Mech. Dyn. Astr.}, 117, 17

\bibitem[Voyatzis (2008)]{gv}
{Voyatzis G.} 2008, \textit{ApJ}, 675, 802


\end{thebibliography}
\end{document}